\documentclass[aps,prc,twocolumn,showpacs,superscriptaddress]{revtex4}  
\usepackage{graphicx}
\usepackage{amsmath}
\begin{document}
\newcommand {\nc} {\newcommand}
\nc {\beq} {\begin{eqnarray}} 
\nc {\eol} {\nonumber \\} 
\nc {\eeq} {\end{eqnarray}} 
\nc {\eeqn} [1] {\label{#1} \end{eqnarray}}   
\nc {\eoln} [1] {\label{#1} \\} 
\nc {\ve} [1] {\mbox{\boldmath $#1$}}
\nc {\rref} [1] {(\ref{#1})} 
\nc {\Eq} [1] {Eq.~(\ref{#1})} 
\nc {\la} {\mbox{$\langle$}}
\nc {\ra} {\mbox{$\rangle$}}
\nc {\dd} {\mbox{${\rm d}$}}
\nc {\cM} {\mathcal{M}} 
\nc {\cY} {\mathcal{Y}} 
\nc {\dem} {\mbox{$\frac{1}{2}$}}
\nc {\ut} {\mbox{$\frac{1}{3}$}} 
\nc {\qt} {\mbox{$\frac{4}{3}$}} 
\nc {\Li} {\mbox{$^6\mathrm{Li}$}}
\nc {\M} {\mbox{$\mathcal{M}$}}

\title{Comparative study of the direct $\alpha+d$ $\rightarrow$ $^6$Li + $\gamma $  astrophysical capture reaction in few-body models}

\author{E.M. Tursunov}
\email{tursune@inp.uz}
\affiliation{Institute of Nuclear Physics,
Academy of Sciences, 100214, Ulugbek, Tashkent, Uzbekistan}

\author{S.A. Turakulov}
\email{turakulov@inp.uz}
\affiliation{Institute of Nuclear Physics,
Academy of Sciences, 100214, Ulugbek, Tashkent, Uzbekistan}

\author{A.S. Kadyrov}
\email{a.kadyrov@curtin.edu.au}
\affiliation{Curtin Institute for Computation and Department of Physics and Astronomy, Curtin University, GPO Box U1987, Perth, WA 6845, Australia}

\begin{abstract}
A comparative analysis of the astrophysical S factor and the reaction rate for the direct  $ \alpha(d,\gamma)^{6}{\rm Li}$ capture reaction, and the primordial abundance of the $^6$Li element, resulting from two-body, three-body and combined cluster models is presented. It is shown that the two-body model, based on the exact-mass prescription, can not correctly describe the dependence of the isospin-forbidden E1 S factor on energy and does not reproduce the temperature dependence of the reaction rate from the direct LUNA data. It is demonstrated that the isospin-forbidden 
E1 astrophysical S factor is very sensitive to the orthogonalization procedure of Pauli-forbidden states  within the three-body model. On the other hand,  
 the E2 S factor does not depend on the orthogonalization method.  
 This insures that the orthogonolizing pseudopotentials method yields a very good description of the  LUNA collaboration's low-energy direct data. At the same time, 
 the SUSY transformation significantly underestimates the data from the LUNA collaboration.
   On the other hand, the energy dependence of the E1 S factor are the same in both methods. The best description of the LUNA data for the  astrophysical S factor and the reaction rates is obtained within the combined E1(three-body OPP)+E2(two-body) model. It yields a value of $(0.72 \pm 0.01) \times 10^{-14}$ for the $^6$Li/H primordial abundance ratio, consistent with the estimation $(0.80 \pm 0.18) \times 10^{-14}$ of the LUNA collaboration.  For the  $^6{\rm Li}/^7{\rm Li}$ abundance ratio an estimation  $(1.40\pm 0.12)\times 10^{-5}$ is obtained in good agreement with the Standard Model prediction.

\end{abstract}

\maketitle

\section{Introduction}

An accurate estimation of the primordial abundance ratio of lithium isotopes still remains one of the  hot topics in nuclear astrophysics.  
Astrophysical data \cite{asp06} yield a value for this  ratio that is three orders of magnitude larger than the big bang nucleosynthesis (BBN) model prediction \cite{serp04}. 
The Nine-Year Wilkinson Microwave Anisotropy Probe (WMAP) collaboration \cite{WMAP} reported an estimated abundance of  $^7$Li/H = $5.13 \times 10^{-10}$ for one of the lithium isotopes. The theory  reproduces this value  \cite{kon13},  however, the $^6$Li/$^7$Li abundance ratio is still not well established. 
The most important input  parameter in the BBN model 
is the rates of the radiative direct 
capture reaction
\begin{eqnarray} \label{1}
\alpha+d\rightarrow {\rm ^6Li}+\gamma
\end{eqnarray}
within the energy range $30 \le E \le 400$ keV \cite{serp04}, where $E$ is the collision energy in the center of mass frame.  
The reaction rates are calculated starting from astrophysical S factors, estimated using  some theoretical model or experimental data.

 After multiple attempts by researchers around the world, first direct experimental results for the astrophysical S factor were obtained by the LUNA collaboration at an underground facility \cite{luna17}. 
The first data set of the LUNA collaboration at energies $E$=94 keV and $E$=134 keV \cite{luna14} was further supplemented at energies $E$=80 keV and $E$=120 keV  \cite{luna17}. As a result, the 
reaction rates estimated on the base of the LUNA data underestimates the old results of the NACRE II collaboration \cite{xu13}. The  BBN model now yields even a lower estimate for the  primordial abundance of the  $^6$Li element. 
On the other hand, fully microscopic calculations of the above process are still not available, although  there are several works devoted to this problem. Some microscopic models deal with the so-called exact-mass prescription for the estimation of the isospin-forbidden E1-transition matrix elements \cite{noll01,TBL91}. Other microscopic studies were limited only to the E2 transition neglecting the important contribution of the E1 transition at low astrophysically-relevant energies. As argued in Ref. \cite{bt18}, the exact-mass prescription which uses the experimental mass values of nuclei,  does not have a misroscopic background and is not applicable to the astrophysical processes like $d(d,\gamma)\alpha$ and $^{12}C(\alpha,\gamma)^{16}O$. 
Hopefully soon, the microscopic models that take into account the isospin mixing will be able to describe the data  \cite{mic18}.

The same exact-mass prescription was used in the two-body potential models  \cite{type97,mukh11,tur15,mukh16,noll01,TBL91,gras17}. In Ref.      \cite{gras17} a detailed comparative analysis was given for the astrophysical S factor of the $^6$Li-formation process using different $\alpha+d$ potential models and corresponding estimations for the $^6$Li abundance were obtained. The two-body models are able to describe the E2 S factor pretty well \cite{tur15} based on the correct asymptotic behaviour of the $^6\rm Li$ bound-state wave function  adjusted to the empirical value  $C_{exp}=2.30 \pm 0.12$ ${\rm fm}^{-1/2}$ of the S-wave asymptotic normalization constant (ANC) \cite{blok93}. However, since the exact mass prescription does not have a microscopic background, it is natural to question how realistic are the results for the isospin-forbidden E1 transition matrix elements, as well as for the cross section and reaction rates, obtained in the two-body models.  

For the solution of the abundance problem of lithium isotopes, the three-body models \cite{tur16,bt18,tur18} are developed beyond the two-body approaches. These models include an important contribution of the  isospin mixing which is responsible for the  isospin-forbidden E1 transition from the initial isosinglet states to the isotriplet components of the $^6$Li ground state. 
Recent study of the isospin-forbidden E1 S factor of the radiative $\alpha$ capture on the  $^{12}$C nucleus within the cluster effective field theory \cite{ando19} confirms this finding. 

 As was shown within the three-body model \cite{bt18,tur18}, at energies below 100 keV the E1 S factor is important, while beyond this region the E2 S factor becomes dominant.  For the calculation of the   three-body $\alpha+p+n$ bound state wave function of the $^6$Li nucleus, a realistic $\alpha-N$ potential \cite{vor95} was used which includes a Pauli-forbidden state in the S wave. In above mentioned works the method of orthogonalising pseudopotentials (OPP) \cite{OPP} has been applied for the elimination of forbidden states in the three-body system. A resulting three-body wave function contains a small isotriplet component of about 0.5 $\%$. That small isotriplet component is responsible for the forbidden  E1 transition and the new  
direct data of the LUNA collaboration for the S factor  
and the reaction rates have been reproduced within the experimental error bars \cite{tur16,bt18,tur18}.  
The estimated $^6$Li/H abundance ratio of $(0.67 \pm 0.01)\times 10^{-14}$  agreed well with the value of $(0.80 \pm 0.18)\times 10^{-14}$, extracted by the LUNA collaboration \cite{luna17}. 
The obtained estimate for the $^6$Li/$^7$Li abundance ratio of $(1.30 \pm 0.12)\times 10^{-5}$  \cite{tur18} is consistent with the standard BBN model \cite{serp04}, and underestimates the astronomical data of $(8.0 \pm 4.4) $\%  \cite{mott17} and of  (0-2)\% \cite{cay07}  by three orders of magnitude. 

On the other hand, a value $C=2.12$ ${\rm fm}^{-1/2}$ of ANC of the S-wave  $\alpha+d$ configuration calculated from the $\alpha+p+n$ three-body bound-state wave function \cite{bt18,tur18} is somewhat smaller than the above mentioned empirical value of $C_{exp}=2.30 \pm 0.12$ ${\rm fm}^{-1/2}$ \cite{blok93} which was reproduced in the two-body model. Owing to the fact that the E2 S factor at low astrophysical energies is mostly defined by the ANC, one can conclude that the three-body model is still not optimal for the estimation of the astrophysical S factor and reaction rates as well as the abundance of the $^6\rm Li$ element.  

The aim of present work is a comparative study of the astrophysical S factor  of the direct  $\alpha(d,\gamma)^{6}{\rm Li}$  capture reaction, as well as the reaction rates and the primordial abundance of the $^6$Li element within few-body cluster models. While the astrophysical S factors and reaction rates have been studied separately in two-body \cite{tur15} and three-body models \cite{tur16,bt18,tur18}, an optimal description of the direct experimental data of the LUNA collaboration has not been obtained. Here we examine a combined  E1(three-body)+E2(two-body) model for the description of the  direct experimental data, based on the viewpoint that the E2 S factor should be better described within the two-body framework due to almost exact reproduction of the empirical value of the ANC. Another interest is to see, how close both absolute values and energy dependence of the E1 and E2 astrophysical S factors in two-body (within the exact-mass prescription) and three-body models to the direct data of the LUNA collaboration. Also important is a behavior of the reaction rate and its temperature dependence.   
Next  important issue is a sensitivity of the results  for the S factor to the projecting method used in the variational calculations of the  $^{6}{\rm Li}$ ground-state wave function. To this end the astrophysical S factor is estimated using the three-body wave function of the $^6$Li ground state, calculated within the supersymmetric transformation (SUSY) method \cite{baye87}. The result are compared with those from the OPP approach \cite{OPP}. 
The OPP method yields a nodal behavior for the $\alpha+N$ relative motion wave function due to a Pauli-forbidden state in the S wave, while the SUSY transformation does not keep this microscopic property, producing a two-body phase-equivalent shallow potential with a core.  

\section{Theoretical model}
 
A formula for the cross section of the radiative capture process can be written in the form
\begin{align}
\sigma_{E}(\lambda)=& \sum_{J_i T_i \pi_i}\sum_{J_f T_f
\pi_f}\sum_{\Omega \lambda}\frac{(2J_f+1)} {\left [I_1
\right]\left[I_2\right]} \frac{32 \pi^2 (\lambda+1)}{\hbar \lambda
\left( \left[ \lambda \right]!! \right)^2} k_{\gamma}^{2 \lambda+1}
\nonumber \\ &\times 
\sum_{l_\omega I_\omega}
 \frac{1}{ k_\omega^2 v_\omega}\mid
 \langle \Psi^{J_f T_f \pi_f}
\|M_\lambda^\Omega\|
\Psi_{l_\omega I_\omega}^{J_i T_i \pi_i}
\rangle \mid^2,
\end{align}
where $\Omega=$E  for electric transition or M for magnetic transition, $\omega$
denotes the entrance channel, $k_{\omega}$, $v_\omega$,  $I_\omega$
are the wave number, velocity of the $\alpha-d$ relative motion and
the spin of the entrance channel, respectively, $J_f$, $T_f$,
$\pi_f$ ($J_i$, $T_i$, $\pi_i$) are the spin, isospin and parity of
the final (initial) state, $I_1$, $I_2$ are channel spins,
$k_{\gamma}=E_\gamma / \hbar c$ is the wave number of the photon
corresponding to the energy $E_\gamma=E_{\rm th}+E$ with the
threshold energy $E_{\rm th}=1.474$ MeV. The wave functions
$\Psi_{l_\omega I_\omega}^{J_i T_i \pi_i} $ and $\Psi^{J_f T_f
\pi_f} $ represent  the initial and final states, respectively.

In the three-body model the initial-state wave function $\Psi_{l_\omega I_\omega}^{J_i T_i \pi_i}$ is factorised into the deuteron wave function and the $\alpha+d$ two-body scattering wave function. The final $\alpha+p+n$ three-body bound-state wave function of the $^6$Li ground state was calculated within the hyperspherical Lagrange-mesh method \cite{ddb03}. 
In the two-body model the initial- and final-state wave functions contain a point-like deuteron. These wave functions are obtained by solving the  two-body Schr\"odinger equation for the $\alpha+d$ bound and scattering states \cite{tur15}, respectively.   
The reduced matrix elements of the E1 and E2 operators are calculated using the initial and final state wave functions described above. Additionally, we use short-hand notations $[I]=2I+1$ and
$[\lambda]!!=(2\lambda+1)!!$.  Details of calculations of the matrix-element can be found in Refs. \cite{tur15,tur16}.

The astrophysical $S$ factor of the process can be written with the help of the cross section as \cite{Fowler}
\begin{eqnarray}
S(E)=E \, \sigma_E(\lambda) \exp(2 \pi \eta),
\end{eqnarray}
where  $\eta$ is the Coulomb parameter.

\section{Numerical results}
 \subsection{Details of calculations}

We performed calculations of the cross section and astrophysical S factor  using the same parameters  as in Refs. \cite{bt18,tur18}. In particular, the
radial wave function of the deuteron is obtained  using the central Minnesota
potential $V_{NN}$ \cite{thom77,RT70} with $\hbar^2/2 m_N=20.7343$ MeV fm$^2$. 
The Schr{\"o}dinger equation for the bound state is solved
using the Lagrange-Laguerre mesh method \cite{baye15}.
It yields $E_d$=-2.202 MeV for the deuteron ground-state energy with
the number of mesh points $N=40$ and a scaling parameter $h_d=0.40$.

The bound and scattering wave functions of the $\alpha-d$ relative motion are
calculated with a deep potential of Dubovichenko \cite{dub94} with a
modification in the $S$ wave \cite{tur15}:
$V_d^{(S)}(R)=-92.44 \exp(- 0.25 R^2) $ MeV. The potential
parameters in the $^3P_0$, $^3P_1$, $^3P_2$  and $^3D_1$, $^3D_2$,
$^3D_3$ partial waves are identical with the Ref. \cite{dub94}. The
potential contains additional Pauli-forbidden states in the $S$ and $P$ waves. The above modification of the
S-wave potential allows one to reproduce the empirical value $C_{\alpha d}=2.31$
fm$^{-1/2}$ of the asymptotic normalization coefficient (ANC) of the
$^6{\rm Li}(1^+)$ ground state which can be extracted from the $\alpha-d$ elastic scattering
cross section \cite{blok93}.

The final three-body $\alpha+p+n$ wave function of the $^6{\rm Li}(1^+)$ ground-state is calculated using
the hyperspherical Lagrange-mesh method \cite{ddb03,tur07,baye15}
with the same Minnesota NN-potential.

The $\alpha-N$ potential of Voronchev {\em et al.}  
\cite{vor95}, containing a deep Pauli-forbidden state in the
$S$-wave was slightly renormalized with the help of a scaling
factor of 1.014 in order to reproduce the bound-state energy of $E_b(^6 {\rm Li})$=3.70 MeV.
The Coulomb potential in the $\alpha - p$ subsystem is taken in the form $2e^2\, \mathrm{erf}(0.83\,R)/R$ \cite{RT70}.
For the solution of the coupled hyperradial equations the Lagrange-mesh method \cite{baye15} was applied. 
A convergence in the energy requires large value of the maximal hypermomentum $K_{\rm max} $  in the hypermomentum expansion. 
It is also important to check  convergence of the isotriplet ($T = 1$) component of
the $^6$Li ground state. 

For the treatment of the Pauli-forbidden states in the three-body system one can use the OPP method \cite{OPP} or the SUSY transformation \cite{baye87} of the initial  $\alpha-N$ nuclear interaction potential.  The first method allows to solve the the three-body Schr{\"o}dinger equation dynamically, while controlling the convergence in the energy with respect to the projecting constant $\lambda$. At large values of $\lambda$ the energy spectrum of the system should not change significantly which means convergence.  The SUSY transformation yields a potential with a core which gives the same phase shift, but one less bound state in the $S$ wave $\alpha-N$ spectrum.  In our previous works we mostly use  the OPP method \cite{bt18,tur18}. First probe of the SUSY method for the astrophysical S factor \cite{TK19} shows a significant difference between the two methods. In the present work we aim to study this comparison in more details including the sensitivity of the reaction rate.  
 
\subsection{Astrophysical S factors}

To start with we note that the two SUSY and OPP methods yield a very similar convergence of the binding energy. In both cases the binding energy of the $^6$Li ground state $E=-3.70$  MeV is obtained with the maximum value of  hypermomentum $K_{max}=24$. However, the isospin structure of the $^6$Li ground-state  wave function in these two cases shows very different pictures.  Specifically, the important isotriplet  ($T=1$)  component of  the $^6$Li ground-state  wave function calculated with the  OPP method has a norm square of about 5.27$\times$10$^{-3}$, while  in the case of the SUSY method the latter is 1.10$\times$10$^{-4}$. 
 As we noted above,  the E1 astrophysical S factor of the $ \alpha(d,\gamma)^{6}{\rm Li}$ direct capture reaction is very sensitive to the isotriplet
 component of the final $^6$Li ground state.  Therefore, the two different orthogonalization methods, SUSY and OPP, should give very  different values for the astrophysical S factor,  
 the rate of the direct capture reaction, and primordial abundance of the $^{6}{\rm Li}$ element. On the other hand, it is important to compare the energy behaviour of the E1 S factor in both cases.

\begin{figure}[htb]
\includegraphics[width=98mm]{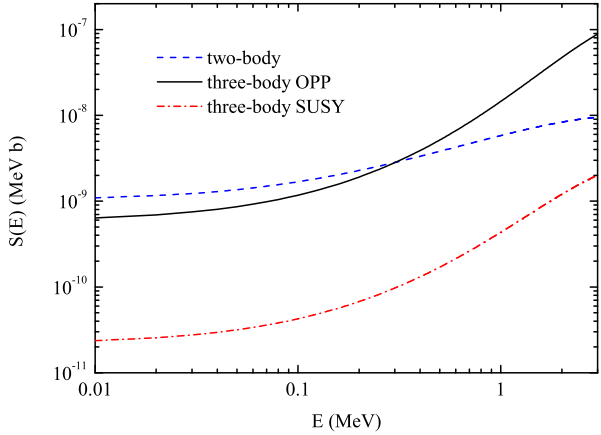}
\caption{Astrophysical E1 S factor  of the direct $ \alpha(d,
\gamma)^{6}{\rm Li}$ capture process. The line for the three-body OPP model is from Refs. \cite{bt18,tur18},
and the line for the two-body model is from Ref. \cite{tur15}.} \label{f1}
\end{figure}

\begin{figure}[htb]
\includegraphics[width=98mm]{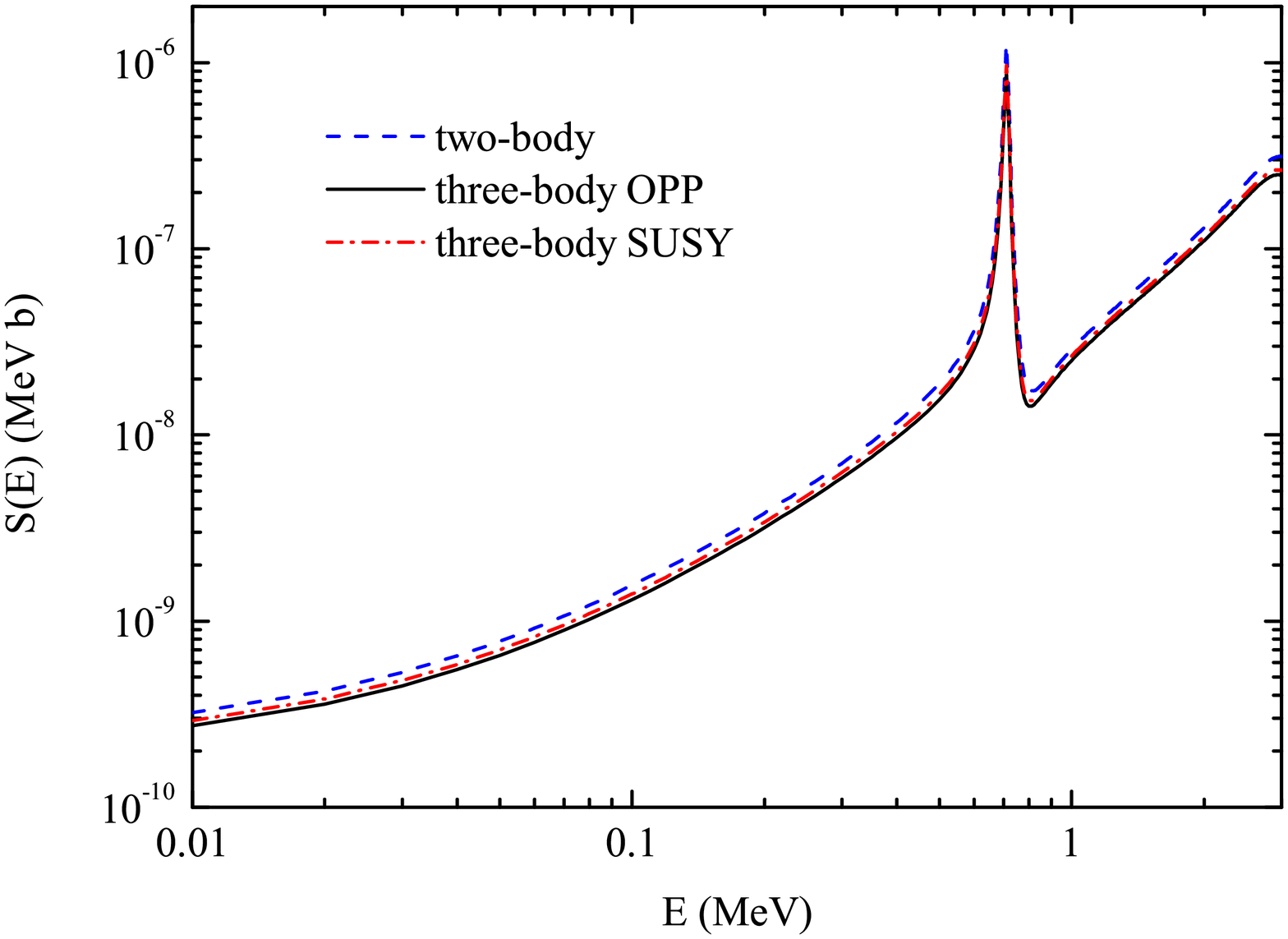}
\caption{Astrophysical E2 S factor  of the direct $ \alpha(d,
\gamma)^{6}{\rm Li}$ capture process. The line for the three-body OPP model is from Refs. \cite{bt18,tur18},
and the line for the two-body model is from Ref. \cite{tur15}.} \label{f2}
\end{figure}

\begin{figure}[htb]
\includegraphics[width=98mm]{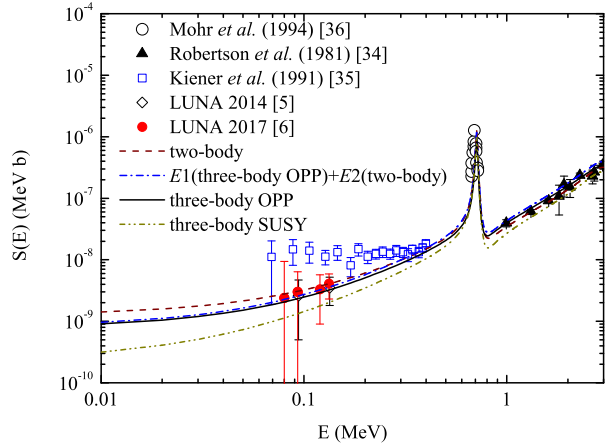}
\caption{Astrophysical S factor  of the direct $ \alpha(d,
\gamma)^{6}{\rm Li}$ capture process. The line for the three-body OPP model is from Refs. \cite{bt18,tur18}, and the line for the two-body model is from Ref. \cite{tur15}. } \label{f3}
\end{figure} 

Figure \ref{f1} displays the E1 astrophysical S factor  of the   $ \alpha(d,
\gamma)^{6}{\rm Li}$ direct radiative capture reaction  
estimated within the two-body model and three-body OPP and SUSY models. As can be seen from the 
figure, the three-body models yield the same behavior but the two-body model gives different energy dependence. On the other hand, 
the S factor, estimated within the SUSY method is more than one order-of-magnitude smaller, than the OPP approach. The obtained result confirms a high sensitive of the E1 astrophysical S factor  to the orthogonalization method when solving the three-body Schr\"odinger equation. Similar effects have been found in Refs. \cite{tur06,tur06a} where the beta-decay transition probability  of the $^6$He halo nucleus to the $\alpha-d$ continuum turns out to be very sensitive to the node position of the S wave $\alpha+d$ scattering wave function at short distances, obtained with the OPP method. At the same time, the SUSY method does not give any nodal behaviour of the scattering or bound-state wave functions at short distances. The same role of the short-range node in the $\alpha+d$ scattering wave function on the M1-transition probability of the $^6 {\rm Li} (0^+)$ isobar-analog state to the $\alpha+d$ continuum was found in Ref. \cite{tur07}. In the present study, the $\alpha-N$ scattering wave function has a short-distance node in the case of using the OPP method which yields  a substantial contribution to the isotriplet 
component  of the total $^6$Li ground-state wave function. The two-body model, as was indicated above, is based on the so-called exact-mass prescription and does not contain any isospin-transition term unlike the three-body models. This is why the two-body model fails to reproduce the energy dependence of the E1 astrophysical S factor.   

The E2 astrophysical S factors calculated  in the two-body and three-body models are 
presented in Fig. \ref{f2}. From the figure one can conclude that the three-body 
OPP and SUSY methods result very similar estimations. 
Thus, one can conclude that the  E2  S factor  is not sensitive to the orthogonalization 
method used in the calculations of the wave function of the $^6$Li ground state. 
On the other hand the two-body E2 S factor  is larger than the three-body result. Indeed, the S-wave ANC of $C_2=2.31$ ${\rm fm}^{-1/2}$ \cite{tur15}  of the $\alpha+d$ configuration within the two-body model is closer to its empirical value of $C_{exp}=2.30 \pm 0.12$ ${\rm fm}^{-1/2}$ \cite{blok93} than the values  $C_3=2.12$ ${\rm fm}^{-1/2}$ (OPP) and $C_3=2.05$ ${\rm fm}^{-1/2}$ (SUSY), calculated from the overlap integral of the $^6$Li ground-state  and deutron wave functions within the three-body models \cite{bt18,tur18}. Since the asymptotic behavior of the bound-state wave function of the $^6$Li nucleus is crucial  for the description of the E2 S factor  at low astrophysical energies, the corresponding two-body results  for the E2 S factor  describe the experimental data better than the three-body models.   

Figure \ref{f3}  shows the theoretical total astrophysical S factors for the reaction of interest. The results are compared with the direct data of the LUNA collaboration \cite{luna14,luna17} and data from Refs. \cite{rob81,kien91,mohr94}.  Owing to the significant  influence of the orthogonalization procedure on the quality of the three-body wave function, one can indeed see a substantial difference between the OPP and SUSY results as the energy reduces. To be specific, the first method describes the direct data of the LUNA collaboration quite well, however, the SUSY transformation leads to a substantial underestimation. On the other hand the two-body model yields larger estimates for the S factor,  especially at low energies although they are still within the experimental error bars of the LUNA data. The best description of the LUNA data is obtained within the combined  E1(three-body OPP)+E2(two-body) model. This combination   
is more preferable, since it is based on the best descriptions of the E1  and E2 S factors within the cluster model. 

\subsection{Reaction rates and abundance of the $^6$Li element}

 For the estimation of the reaction rates we use the formalism which was described in details in Ref.\cite{tur18}. When  variable $k_{\text{B}}T$ is expressed in units of MeV, where $k_{\text{B}}$ is the Boltzmann coefficient, it is
customary to use a new variable $T_9$ for the temperature in units of
$10^9$ K according to the equation $k_{\text{B}}T=T_{9}/11.605$ MeV.
In our calculations $T_9$ varies within the interval $0.001\leq T_{9}
\leq 10$.

\begin{figure}[htb]
\includegraphics[width=98mm]{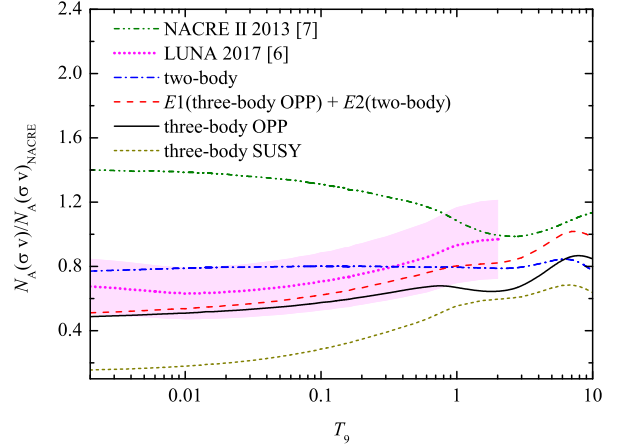}
\caption{Reaction rates of the direct $\alpha+d\rightarrow
^6$Li$+\gamma$ capture process within the two-body and three-body
models normalized to the NACRE 1999 experimental data \cite{ang99}. The line for the three-body OPP model is from Ref. \cite{tur18}. The shaded area is for the error bar of the LUNA 2017 data \cite{luna17}.} \label{f4}
\end{figure}

In Fig. \ref{f4} we display the theoretical reaction rates of the  direct   $ \alpha(d,\gamma)^{6}{\rm Li}$ capture process calculated within two-body and three-body models in the temperature interval   $10^{6}$ K $\leq T \leq 10^{10}$ K ($ 0.001\leq T_{9} \leq 10 $), in comparison with the LUNA 2017 analysis \cite{luna17} and the results of the NACRE II collaboration \cite{xu13}, normalized to the standard NACRE 1999 data \cite{ang99}. As can be seen from the figure, the three-body OPP method yields a good description of the direct LUNA data, while the three-body SUSY model  goes much below the error bar. In addition, the OPP model reproduces the temperature dependence of the reaction rates. Although the two-body model results lie within the experimental error bars of the LUNA data, the temperature dependence of the data is not reproduced.  Again, the best description of the direct LUNA data for the absolute values and temperature dependence of the reaction rates is obtained within the combined E1(three-body OPP)+E2(two-body) model.   

For the estimation of the abundance of the $^6$Li element, the theoretical reaction rate is approximated within 1.40\% (the two-body model), 1.89\% (the three-body OPP model) and  1.88\% (the combined model) by the following analytical formula:
\begin{align}
 N_{24}(\sigma v)=&p_0 T_{9}^{-2/3} \exp ( -7.423 T_{9}^{-1/3}) \times ( 1 + p_1 T_9^{1/3} 
 \nonumber \\ 
 &+ p_2 T_{9}^{2/3}+p_3 T_{9}+p_4 T_{9}^{4/3}  
+ p_5 T_{9}^{5/3}
+p_6 T_{9}^2
\nonumber \\ 
&+ p_7 T_{9}^{7/3}) + p_8 T_{9}^{-3/2} \exp (-7.889 T_9^{-1}).
\end{align}
The coefficients of the analytical polynomial
approximation of the $d(\alpha,\gamma)^{6}$Li reaction rates
estimated within the two-body and three-body OPP models 
are given in Table 1 in the temperature interval $0.001\leq T_{9} \leq 10 $.

\begin{table*}[htb]
\caption {Fitted values of the coefficients of analytical
approximation for the reaction rates of the direct capture process $\alpha({\rm d},
\gamma)^{6}{\rm Li}$}
\label{tab02}
\begin{center}
{\scriptsize

\begin{tabular}{|c|c|c|c|c|c|c|c|c|c|}
\hline \textrm{Model} & $p_0$ & $p_1$ & $p_2$ & $p_3$ & $p_4$ & $p_5$ & $p_6$ & $p_7$ & $p_8$ \\
\hline
\textrm{two-body} &9.659 &-2.223 &29.296 &-96.733& 169.841&-140.218 & 57.705 & -9.152&61.663\\
\textrm{three-body OPP} &6.004 &-2.558 &34.730 &-115.482& 205.801&-169.456 & 71.428 & -11.614&42.354 \\
\textrm{ E1(three-body OPP)+E2(two-body)} &6.741 &-3.731 &41.646 &-137.035& 242.156&-201.509 & 85.095 & -13.814&61.752 \\
\hline
\end{tabular}
}
\end{center}
\end{table*}

On the basis of the theoretical reaction rates and with the help of the publicly available 
PArthENoPE \cite{Pisanti08} code we have estimated the
primordial abundance of the $^6$Li element. If we adopt the Planck
2015 best fit for the baryon density parameter
  $\Omega_b h^2=0.02229^{+0.00029}_{-0.00027}$  \cite{ade16}   and the neutron lifetime $\tau_n=880.3 \pm
1.1$ s  \cite{olive14}, for the $^6$Li/H abundance ratio we have an estimation of (0.89-0.92)$\times 10^{-14}$,  (0.66 - 0.68)$\times 10^{-14}$, and (0.71-73)$\times 10^{-14}$ within the two-body, three-body OPP,  and combined E1(three-body OPP)+E2(two-body) models, respectively.
These numbers are consistent with the new estimation $^6$Li/H=$(0.80 \pm 0.18) \times 10^{-14}$ of the LUNA collaboration \cite{luna17}.
However, as shown above, the two-body model does not reproduce the temperature dependence of the experimental reaction rates, hence the corresponding estimation for the $^6$Li/H abundance ratio is not realistic. If we adopt a value of $(5.2\pm 0.4) \times 10^{-10}$ \cite{kon13} for the $^7{\rm Li}/{\rm H}$ abundance ratio, then within the combined model we have an estimation of $(1.40\pm 0.12) \times 10^{-5}$ for the $^6{\rm Li}/^7{\rm Li}$ abundance ratio. The latter agrees well with the Standard Model estimation \cite{serp04}.

\section{Conclusion}
The astrophysical S factor  and the reaction rates of the direct  $ \alpha(d,\gamma)^{6}{\rm Li}$ capture reaction, as well as the primordial abundance of the $^6$Li element have been estimated within two-body, three-body and combined cluster models. It is shown that although the two-body model based on the exact mass prescription can describe the astrophysical S factor  of the LUNA collaboration within the experimental error bars, at the same time, it does not reproduce the temperature dependence of the reaction rate. This is a consequence of the fact the exact-mass prescription for the estimation of the isospin-forbidden E1 transition matrix elements is invalid.  Within the three-body model a sensitivity of the theoretical astrophysical S factor  to the orthogonalization procedure has been examined. It is found that the isospin-forbidden E1 S factor  is very sensitive to the orthogonalization procedure used during the calculations of the $^6{\rm Li}$ ground state wave function. However, the E2 S factor  shows a different picture and is independent of the orthogonalization method. As a conclusion, the OPP method describes the direct LUNA data very well, while the SUSY transformation significantly underestimates them. However, one can note that both methods yield the same energy dependence of the E1 S factor. 

The best description of the LUNA data for the  astrophysical S factor and the reaction rates is obtained within the combined E1(three-body OPP)+E2(two-body) model. It yields a value of $(0.72 \pm 0.01) \times 10^{-14}$ for the $^6$Li/H primordial abundance ratio, consistent with the estimation of $(0.80 \pm 0.18) \times 10^{-14}$ of the LUNA collaboration. And for the  $^6{\rm Li}/^7{\rm Li}$ abundance ratio an estimation  of $(1.40\pm 0.12) \times 10^{-5}$ was obtained in agreement  with the Standard Model prediction.

\acknowledgments
The authors thank D. Baye and P. Descouvemont for useful discussions of the presented results.
ASK was supported by the Australian Research Council.


\begin{thebibliography}{00}    
\bibitem {asp06} M.~Asplund, {\it et~al. Astrophys. J.},  {\bf 644}, 229 (2006).
\bibitem {serp04} P.D.~Serpico, {\it et~al.}, {\it J. Cosmol. Astropart. Phys.} {\bf 12}, 010 (2004).
\bibitem{WMAP} G.~Hinshaw {\it et al.}, {\it Astrophys. J. Suppl.} {\bf 208} 19 (2013).
\bibitem{kon13} A.~Kontos, E.~Uberseder, R.~de~Boer {\it et al.}, {\it Phys. Rev.} {\bf C87} 065804 (2013). 
\bibitem {luna17} LUNA Collaboration (D.~Trezzi, {\it et~al.}) {\it  Astropart. Phys.} {\bf 89} 57 (2017).
\bibitem {luna14} LUNA Collaboration (M.~Anders, {\it et~al.}) {\it  Phys.~ Rev.~ Lett.}  {\bf 113}, 042501 (2014).
\bibitem{xu13} Y. Xu, K. Takahashi, S. Goriely, {\it et~al.} (NACRE II), {\it Nuclear Physics} {\bf  A918} 61 (2013).
\bibitem{noll01}  K.M.~Nollett, R.B.~Wiringa and R.~Schiavilla, {\it  Phys. Rev. C}  {\bf 63} 024003 (2001).
\bibitem{TBL91} S.~Typel, G.~Bl\"uge and K.~Langanke, {\it Z. Phys. A} {\bf 339} 335 (1991).
\bibitem{bt18} D. Baye and E. M. Tursunov, {\it J. Phys. G: Nucl. Part. Phys.} {\bf 45}  085102 (2018).
\bibitem{mic18}S.~Mickevicius, A.~Stepsys, D.~Germanas, and R.K.~Kalinauskas,
{\it Phys. Atom. Nuclei} {\bf 81} 899 (2018).
\bibitem{type97} S.~Typel, H.~Wolter and G.~Baur,  {\it Nucl. Phys. A} {\bf 613} 147 (1997).
\bibitem{mukh11} A.M.~Mukhamedzhanov, L.D.~Blokhintsev and B.F.~Irgaziev, {\it  Phys. Rev. C} {\bf 83} 055805 (2011).
\bibitem{tur15} E.M.~Tursunov, S.A.~Turakulov and P.~ Descouvemont, {\it Phys. At. Nucl.} {\bf 78} 193 (2015).
\bibitem{mukh16} A.M.~Mukhamedzhanov, Shubhchintak and C.A.~Bertulani, {\it Phys. Rev.} {\bf    C93} 045805 (2016).
\bibitem{gras17}A.~Grassi, G.~Mangano, L.E.~Marcucci, and O.~Pisanti, {\it Phys. Rev. C} {\bf 96} 045807 (2017).
\bibitem{blok93} L.D.~Blokhintsev {\it et~al.} {\it Phys. Rev. C} {\bf 48} 2390 (1993).
\bibitem{tur16} E. M. Tursunov, A. S. Kadyrov, S. A. Turakulov and I. Bray, 
{\it  Phys. Rev. C}  {\bf 94} 015801 (2016).
\bibitem{tur18} E. M. Tursunov,  S. A. Turakulov, A. S. Kadyrov and I. Bray, 
{\it  Phys. Rev. C}  {\bf 98} 055803 (2018).
\bibitem{ando19} Shung-Ichi~Ando, {\it Phys. Rev. C} {\bf 100}, 015807 (2019).
\bibitem{vor95} V.T.~Voronchev, V.I.~Kukulin, V.N.~Pomerantsev  and G.~Ryzhikh, {\it Few-Body Syst.} {\bf 18} 191 (1995).
\bibitem{OPP}V.I.~Kukulin, V.N.~Pomerantsev  and E.M.~Tursunov, {\it Phys. At. Nucl.} {\bf 59} 757 (1996).
\bibitem{mott17} A.~Mott, M.~Steffen, E.~Caffau, F.~Spada  and K.~G.~Strassmeier, {\it Astron. Astrophys.} {\bf 604} A44  (2017).  
\bibitem{cay07}  R.~Cayrel, M.~Steen, H.~Chand {\it et al.}, {\it Astron. Astrophys.} {\bf 473}  L37 (2007). 
 \bibitem{baye87} D.~Baye, {\it  Phys. Rev. Lett. } {\bf 58} 2738 (1987). 
\bibitem{ddb03} P.~Descouvemont, C.~Daniel, and D.~Baye,{\it  Phys. Rev. C} {\bf 67} 044309 (2003).
\bibitem{Fowler} W.A.~ Fowler, G.R.~Gaughlan and B.A.~Zimmerman, {\it Ann. Rev. Astronom.  Astrophys.} {\bf  13} 69 (1975).
\bibitem{thom77} D.~Thompson, M.~Lemere and Y.~Tang, {\it Nucl. Phys. A} {\bf 286} 53 (1977).
\bibitem{RT70} I.~Reichstein and Y. C.~Tang, { \it Nucl. Phys. A} {\bf 158}  529 (1970).
\bibitem{baye15} D.~Baye, {\it Phys. Rep.} {\bf 565} 1 (2015). 
\bibitem{dub94} S.B.~Dubovichenko  and A.V.~Dzhazairov-Kakhramanov, {\it Phys. At. Nucl.} {\bf 57} 733 (1994).
\bibitem{tur07} E. M.~Tursunov, P.~Descouvemont and D.~Baye, Nucl.
Phys. A {\bf 793} 52 (2007).
\bibitem{TK19} E. M. Tursunov and A. S. Kadyrov. Int. J. Mod. Phys: CS  49, 1960015 (2019)
\bibitem{tur06}  E.M.~Tursunov, D.~Baye and P.~ Descouvemont,  {\it Phys. Rev. C} {\bf 73} 014303 (2006).
\bibitem{tur06a}  E.M.~Tursunov, D.~Baye and P.~ Descouvemont,  {\it Phys. Rev. C} {\bf 74} 069904 (2006).
\bibitem{rob81}R.G.H.~Robertson {\it et~al.} {\it Phys. Rev. Lett.} {\bf 47} 1867 (1981).
\bibitem{kien91} J.~Kiener {\it et~al.} {\it Phys. Rev. C} {\bf 44} 2195 (1991).
\bibitem{mohr94} P.~Mohr {\it et~al.} {\it Phys. Rev. C} {\bf 50} 1543 (1994).
\bibitem{ang99} C. Angulo {\it et~al.} (NACRE), {\it Nuclear Physics} {\bf  A656} 3 (1999).
\bibitem{Pisanti08} O.~Pisanti, A.~Cirillo, S.~Esposito, F.~Iocco, G.~Mangano,
G.~Miele, and P.D.~Serpico, Comput. Phys. Commun. {\bf 178} 956 (2008).
\bibitem{ade16}P.A.R.~Ade et al. (Planck Collaboration), Astron. Astrophys.
{\bf 594} A13 (2016) 
\bibitem{olive14} K.A.~Olive, K.~Agashe, C.~Amsler, et al.Chin. Phys., {\bf C38}  090001 (2014).
\end{thebibliography}
\end{document}